\documentclass[final]{svjour2}
\usepackage{graphicx}
\usepackage{rotating}
\usepackage{amssymb}
\usepackage{mathptmx}
\usepackage[numbers]{natbib}
\usepackage{fancyhdr}
\makeatletter
\journalname{Journal of Low Temperature Physics}

\bibpunct{}{}{,}{s}{}{,}
\begin{document}

\newcommand{\hdblarrow}{H\makebox[0.9ex][l]{$\downdownarrows$}-}
\title{First polarised light with the NIKA camera}
\author{A. Ritacco$^1$ \and R. Adam$^1$ \and A. Adane$^2$\and P. Ade$^3$\and P. Andr\'e$^4$\and A. Beelen$^5$\and B. Belier$^6$\and A. Beno\^it$^7$\and 
A. Bideaud$^3$\and N. Billot$^8$\and O. Bourrion$^1$\and M. Calvo$^7$\and A. Catalano$^1$\and G. Coiffard$^2$\and B. Comis$^1$\and  A. D'Addabbo$^{7,9}$\and F.-X. D\'esert$^{10}$\and S. Doyle$^3$\and J. Goupy$^7$\and C. Kramer$^8$\and S. Leclercq$^2$\and J.F. Mac\'ias-P\'erez$^1$ \and J. Martino$^5$\and P. Mauskopf$^{3,12}$\and A. Maury$^4$\and F. Mayet$^1$\and A. Monfardini$^7$\and F. Pajot$^5$\and E. Pascale$^3$\and L. Perotto$^1$\and G. Pisano$^3$ \and N. Ponthieu$^{10}$\and M. Rebolo-Iglesias$^1$\and V. Rev\'eret$^4$\and L. Rodriguez$^4$\and G. Savini$^{11}$\and K. Schuster$^2$\and A. Sievers$^8$\and C. Thum$^{2}$\and  S. Triqueneaux$^7$\and C. Tucker$^3$\and R. Zylka$^2$}

\institute{$^1$  LPSC, Universit\'e Grenoble-Alpes, CNRS/IN2P3, 53, rue des Martyrs, Grenoble, France \\
$^2$ Institut de RadioAstronomie Millim\'etrique (IRAM), Grenoble, France \\
$^3$ Astronomy Instrumentation Group, University of Cardiff, UK \\
$^4$ Laboratoire AIM, CEA/IRFU, CNRS/INSU, Universit\'e Paris Diderot, CEA-Saclay, France \\
$^5$ Institut d'Astrophysique Spatiale (IAS), CNRS and Universit\'e Paris Sud, Orsay, France \\
$^6$ Institut d'Electronique Fondamentale (IEF), Universit\'e Paris Sud, Orsay, France \\
$^7$ Institut N\'eel, CNRS and Universit\'e de Grenoble, France \\
$^8$ Institut de RadioAstronomie Millim\'etrique (IRAM), Granada, Spain \\
$^9$ Dipartimento di Fisica, La Sapienza Universit\`a di Roma, Italy \\
$^{10}$ IPAG, CNRS and Universit\'e de Grenoble, France \\
$^{11}$ University College London, Department  of Physics and Astronomy, London, UK\\
$^{12}$ School of Earth and Space Exploration and Department of Physics, Arizona State University
\email{alessia.ritacco@lpsc.in2p3.fr}}
\maketitle
\begin{abstract}NIKA is a dual-band camera operating with  315 frequency multiplexed LEKIDs cooled at 100 mK. NIKA is designed to observe the sky in intensity and polarisation at 150 and 260 GHz from the IRAM 30-m telescope. It is a test-bench for the final NIKA2 camera. The incoming linear polarisation is modulated at four times the mechanical rotation frequency by a warm rotating multi-layer Half Wave Plate. Then, the signal is analysed by a wire grid and finally absorbed by the Lumped Element Kinetic Inductance Detectors (LEKIDs). The small time constant ($<$ 1ms ) of the LEKID detectors combined with the modulation of the HWP enables the quasi-simultaneous measurement of the three Stokes parameters I, Q, U, representing linear polarisation. In this paper we present results of recent observational campaigns demonstrating the good performance of NIKA in detecting polarisation at mm wavelength.
\end{abstract}

\section{Introduction}
High-resolution millimeter-wavelength photometry and polarimetry are needed for mapping the magnetic field within Galactic regions \cite{planck}.The polarisation degree of continuum emission is low (typically $< 5 $ $ \% $ \cite{matthew}), a polarimetric study of nearby star-forming filaments and pre-stellar cores thus requires large improvement in sensitivity which can be achieved with the mapping speed of NIKA and foreseen for NIKA2. NIKA2 will have about 5000 detectors,  $\sim$ 1000 at 150 GHz (2.05 mm),  $\sim$ 4000 with polarisation capability at 260 GHz (1.15 mm), a field of view (FOV) of 6.5' and will be installed at the end of 2015 at the IRAM 30-m telescope (Pico-Veleta, Granada, Spain). The prototype NIKA camera, with  reduced number of detectors and FOV has demonstrated the scientific quality of LEKID detectors via intensity observations of various astrophysical sources \cite{catalano} and in particular of the Sunyaev-Zel'dovich effect on clusters of galaxies \cite{adam,adam2}. In this work, we focus on the description of the NIKA polarisation capabilities. In section 2 we describe the NIKA instrumental setup including polarisation facilities.  Section 3 presents the polarisation tests carried out in laboratory. Section 4 and 5 discuss the observation strategy and data analysis. Finally, in section 6 we present the NIKA first lights in  polarisation and related results. 

\section{NIKA instrument}
NIKA is composed of two LEKID (Lumped Element Kinetic Inductance Detector) arrays of hundreds of pixels, both cooled down to 100 mK by a $^3$He-$^4$He dilution cryostat. The optical coupling between the telescope and the detectors is made by warm aluminium mirrors and cold refractive optics \cite{catalano}. The NIKA detector design consists of a 3$^{rd}$ order Hilbert fractal curve \cite{roesch} to allow the absorption of both polarisation orientations. An optical study has been performed to check the absorption  in  the  two  polarisations and the results  show  the same  sensitivity  for  both  polarisations \cite{roesch}. The NIKA polarisation equipment consists of a multi-layer half wave plate (HWP), a motor to rotate it and a fixed polariser to select the direction of polarisation. All the system is at room temperature and placed on a warm pupil at {$\sim$} 10 cm from the cryostat window.

\section{Laboratory characterisation of the HWP efficiency}
A typical mono-chromatic HWP consists of a birefringent material where two orthogonal directions, defining the ordinary and extraordinary axes, have two different optical indeces $n_o$ and $n_e$.  The ordinary and extraordinary components of an incident electric field  passing through the HWP are thus phase-shifted. 
Given the operating wavelength {$\lambda_0$}, the phase shift is achieved by tuning the HWP thickness $d$ that corresponds to the optical path inside the element. In the generic expression of Mueller matrix for an HWP \cite{savini} {$\alpha$} and {$\beta$} are the transmission coefficients of the orthogonal polarisations aligned with the birefringent axes and   {$\phi$} is the phase shift introduced by the plate between the orthogonal polarisations. 
In an ideal case {$\alpha$} and {$\beta$} are equal to 1 and  {$\phi$} = {$\pi$}. Ideally, in the continuous rotation mode a linearly polarised signal is modulated by the combined action of the HWP and subsequent polariser at four times the mechanical rotation frequency {$\omega$}. In this case a KID $k$ measures a signal described by the Stokes vectors $I$, $Q$, $U$:
\begin{equation}
 S_{k} =  \frac{1}{2}\{I + {\rho}_{\rm pol}[Q\cos(4{\omega}t) +  U\sin({4{\omega}t})]\}
 \label{signal_polar}
 \end{equation}
 where ${\rho}_{\rm pol}$ is the polarisation efficiency. An intrinsic limit in designing a HWP totally transmitting in a broad spectral range is represented by the dependence of the phase-shift on wavelength.
 For the broad spectral range of NIKA, we chose a multi-layer HWP \cite{savini}. 
 The polariser is necessary to select the direction of the polarisation.

In the following, we assume an ideal polariser and we focus only on the characterisation of the HWP parameters: the transmissions {$\alpha$} and {$\beta$}, the polarisation efficiency $\rho_{\rm pol}$. The HWP, manufactured at the Cardiff University, was characterised to reach the maximum transmission in the NIKA range frequency.
\begin{figure}
\begin{center}
\includegraphics[%
	width=0.65\linewidth,
  keepaspectratio]{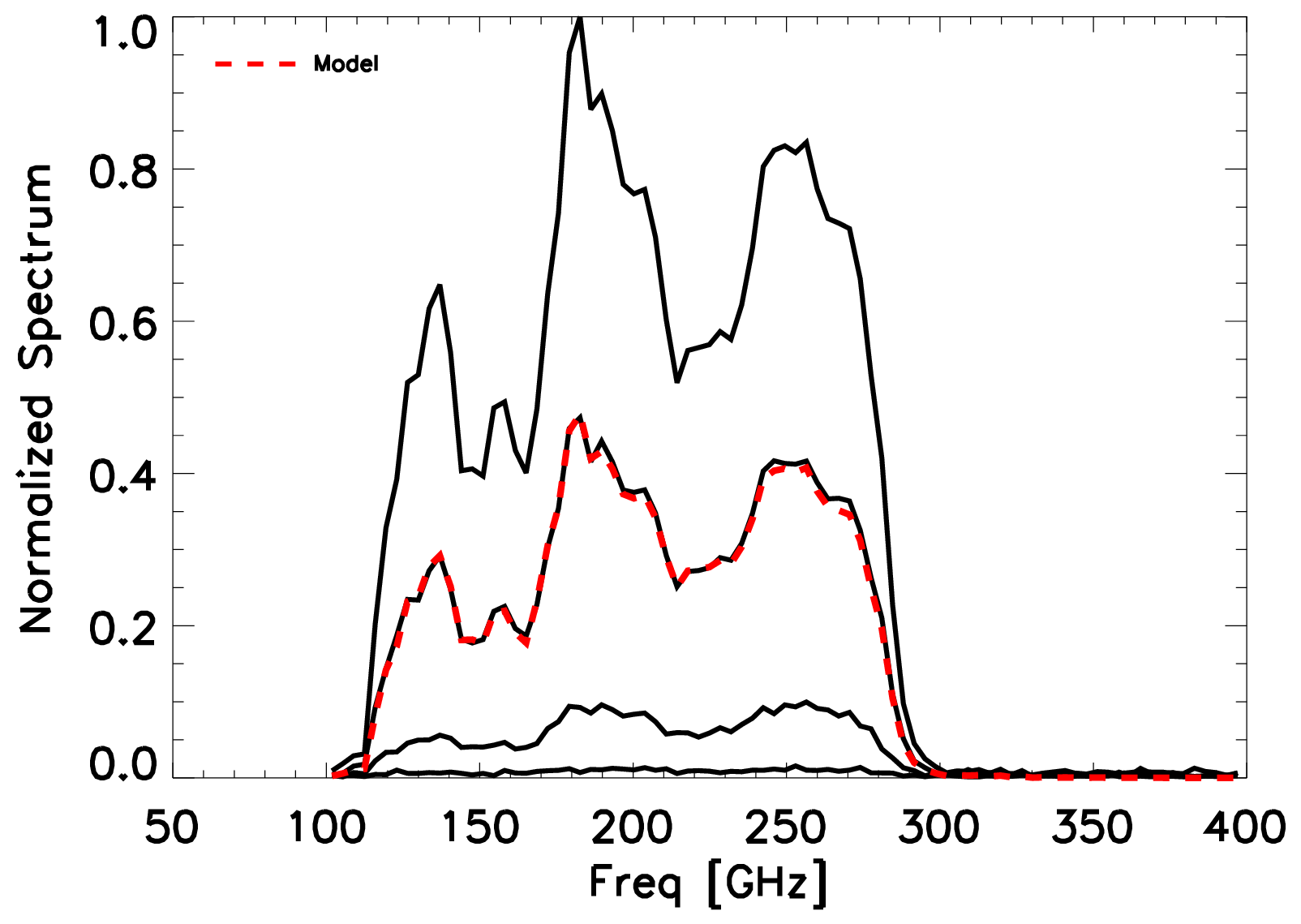}
  \caption{\footnotesize Maximum transmission at an angle of 46.8$^\circ$ respect to the HWP zero and attenuated spectra at 72$^\circ$, 79$^\circ$, 86.4$^\circ$ (top curve to bottom curve). For example, the model (red dotted line) fits the spectrum at an angle of 72$^\circ$.}
				\label{fig:spectre}	
\end{center}
\end{figure}

\begin{figure}
\begin{center}
\includegraphics[%
	width=0.65\linewidth,
  keepaspectratio]{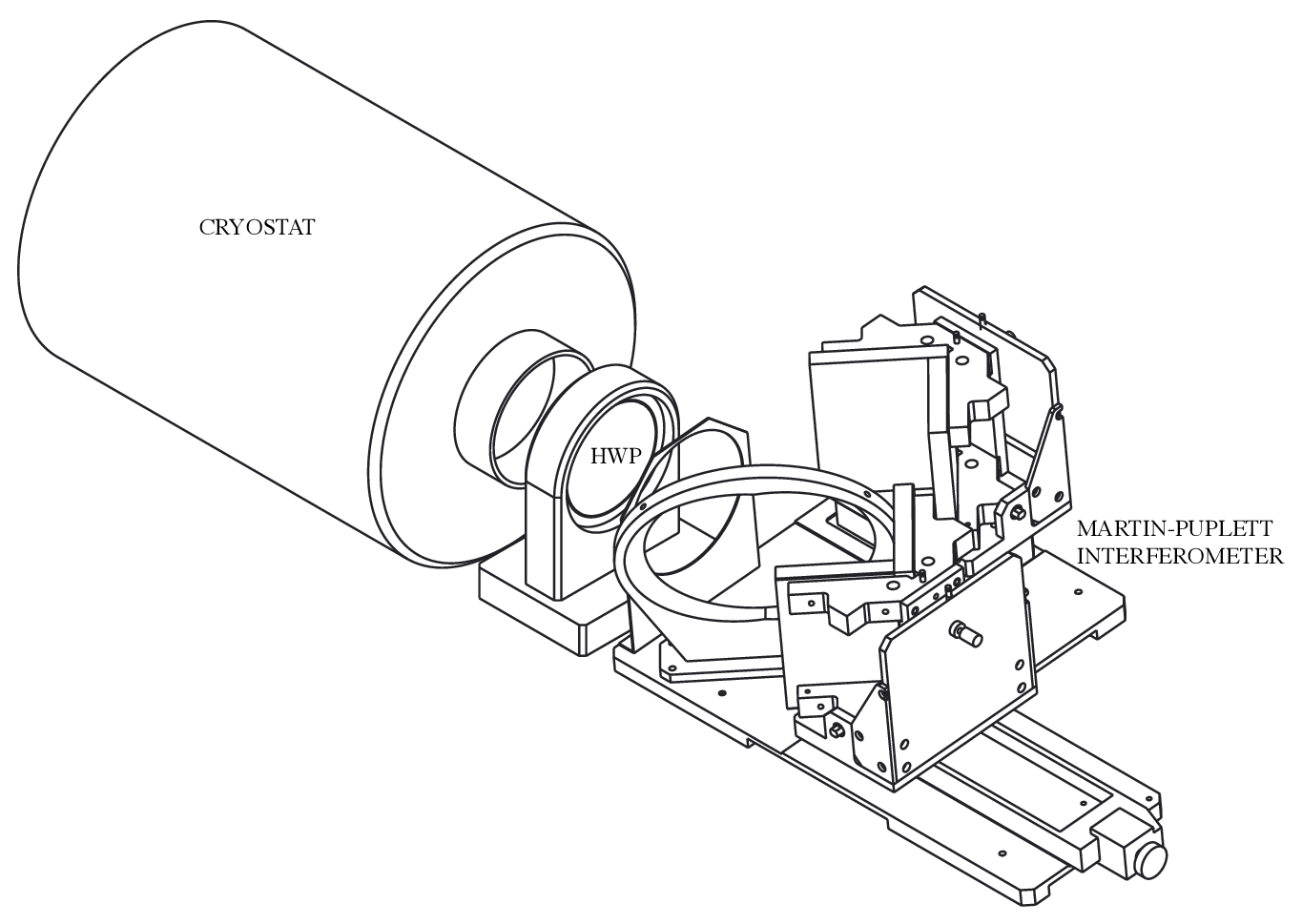}
  \caption{\footnotesize Laboratory instrumental setup. From left to right, the cryostat used to cool down the KIDs to 100 mK, the HWP and polarimeter, and the Martin Puplett interferometer.}
				\label{fig:setup}
\end{center}
\end{figure}

In order to estimate the performance of the whole polarisation chain and the contribution from any instrumental polarisation we have performed the laboratory characterisation. We used a polarising Martin Puplett type Fourier-Transform-Spectrometer, the HWP in a fixed position and the dilution cryostat that cools down the optics and KIDs to 100 mK, see the fig. ~ \ref{fig:setup}. In fig.~\ref{fig:spectre} the maximum transmission is represented by the spectrum at about 46.8$^\circ$ with respect to the HWP zero. Rotating the HWP we found an attenuation as shown in fig.~\ref{fig:spectre}. Creating a model accounting for a realistic HWP Mueller matrix \cite{savini} we can derive the HWP parameters.
In this model we use the phase shift $\phi$ optimised at each wavelength of the NIKA band-pass provided by Cardiff University. Fitting the observed attenuated spectra, we find the values $\alpha$ and $\beta$ of 1 and 0.98 at 1.15 mm and 0.98 and 0.92 at 2.05 mm.
Once the $\alpha$, $\beta$ and $\phi$ parameters are determined we can derive the polarisation efficiency $\rho_{\rm pol}$. We find $\rho_{\rm pol}$ $\simeq$ 1 almost constant in the NIKA band, so we observe a transmission of 100 \% in polarisation.

\section{Observation strategy and data processing pipeline}
In order to obtain quasi-simultaneous measurements of the three Stokes parameters (I, Q, U) on a given sky position, we chose to rotate the HWP continuously. This rotation is fast ({$\omega$} $\sim$ 3 Hz) compared to the scan speed (max 35 arcsec/s) and the beam width (larger than 12'' FWHM). This provides 4 and 6 independent measurements of I, Q, U per FWHM at 1.15 mm and at 2.05 mm, respectively.
The output polarised data are expected to be modulated sinusoidally at four times the rotation frequency {$\omega$} of the HWP. Thus, the signal is extracted using a lock-in procedure around the fourth harmonic of this frequency. In real conditions there is an additional parasitic signal at harmonics of 1{$\omega$}, 2{$\omega$}, 3{$\omega$} due to imperfections of the HWP. 
Accounting for these parasitic signals eq. (\ref{signal_polar}) reads:
\begin{eqnarray}
 S_{k} &=& \frac{1}{2}[I + {\rho}_{\rm pol}Q\cos(4{\omega}t + 2{\alpha}_{\rm Sky}(p(t))) +  {\rho}_{\rm pol}Usin({4{\omega}t} + 2{\alpha}_{Sky}(p(t)))  \nonumber \\
          &+& S_{\rm parasitic} ({\omega}t, 2{\omega}t, 3{\omega}t, ...)] + {\rm atmosphere} + {\rm noise}_{\rm detector}
 \label{pol_eq}
 \end{eqnarray}
 where $p(t)$ represents the pointing matrix and ${\alpha}_{Sky}$ the angle between the telescope reference frame and the local meridian on the sky. 
The angle ${\alpha}_{\rm Sky}$ is measured from north to east in the equatorial system as  ${\alpha}_{\rm Sky} = {\tau} - {\epsilon} - {\eta}$,
where ${\epsilon}$ represents the elevation, ${\eta}$ the parallactic angle, and ${\tau}$ = 45.54$^{\circ}$ is the tilt of the switch mirror of the Nasmyth system along the elevation axis, which creates an angular offset of the image of the sky on the detectors arrays. Atmospheric emission dominates at low frequencies with a 1/f like spectrum. Combining data from multiple detectors we remove atmospheric common modes from the signal. The modulation of the polarised signal imposed by the HWP combined with the scanning strategy allows us to shift the astrophysical signal away from the largest atmospheric contribution \cite{johnson}. The accuracy of this strategy lies in the calibration of the HWP and the validity of the data analysis pipeline. 
We have developed a dedicated polarisation pipeline based on the intensity pipeline presented in \cite{adam,catalano}. I, Q, U Stokes parameters maps in sky coordinates are constructed by applying the following procedures:
i) construction of a template of the HWP modulation and removal of parasitic signal;
ii) construction of Q and U TOIs (time ordered informations) from the 4{$\omega$}t component;
iii) removal of atmospheric noise both in intensity and polarisation;
iv) and, projection of the I, Q, U TOIs into I, Q, U maps.
\section{First lights at the telescope}
For the polarisation angle, the HWP zero is defined with respect to the Nasmyth coordinates reference frame. A synchronised acquisition software provides the rotation angles of the HWP.  The mechanical modulator completes 100 steps to cover 360{$^\circ$}. Thus, our precision in the determination of the HWP zero is $\sim$ 1.8{$^\circ$}. 
\subsection{Systematic effects and first polarised results}
\begin{figure}
         \includegraphics[%
         	width=0.33\linewidth,keepaspectratio]{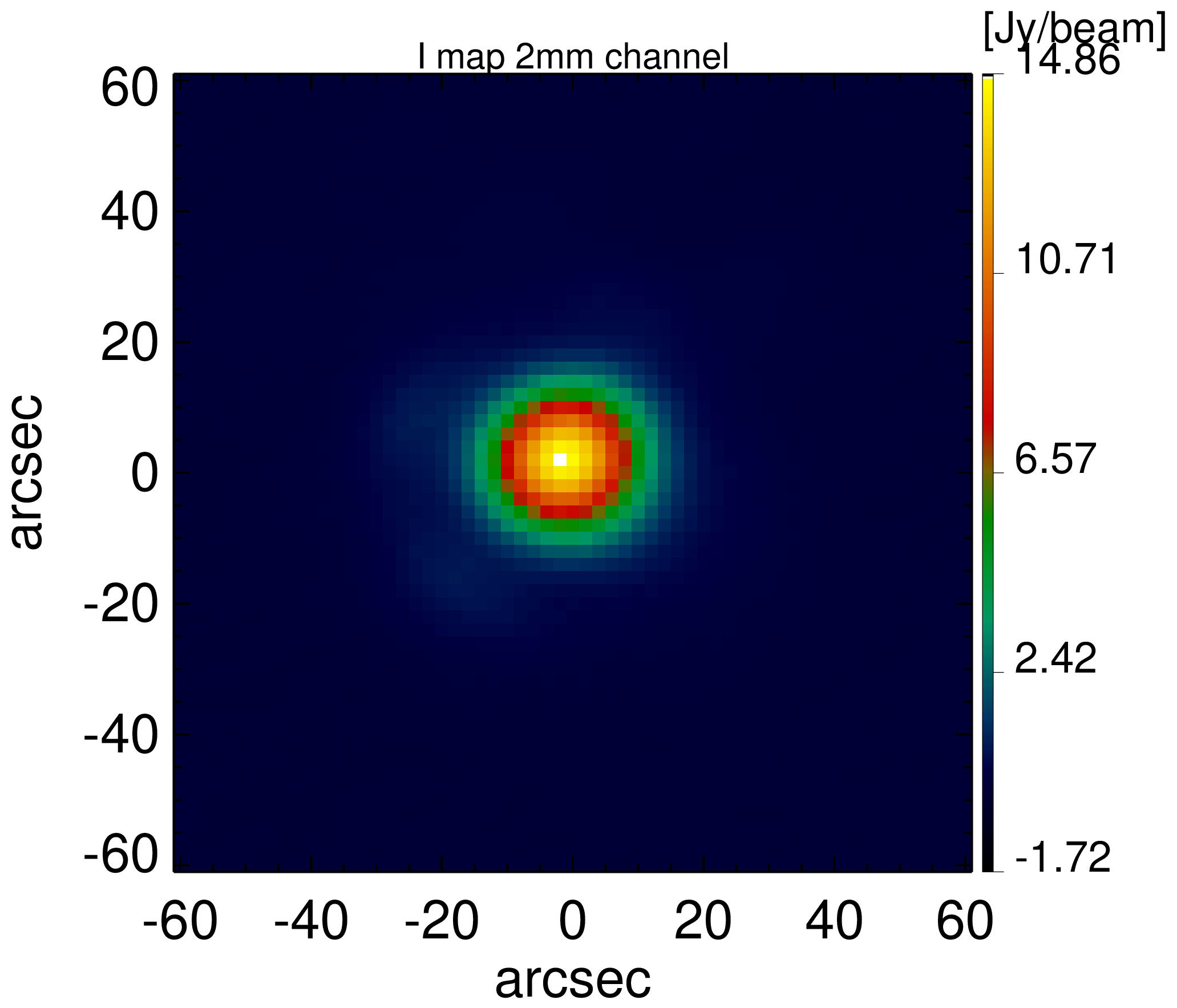}
         \includegraphics[%
         	width=0.33\linewidth,keepaspectratio]{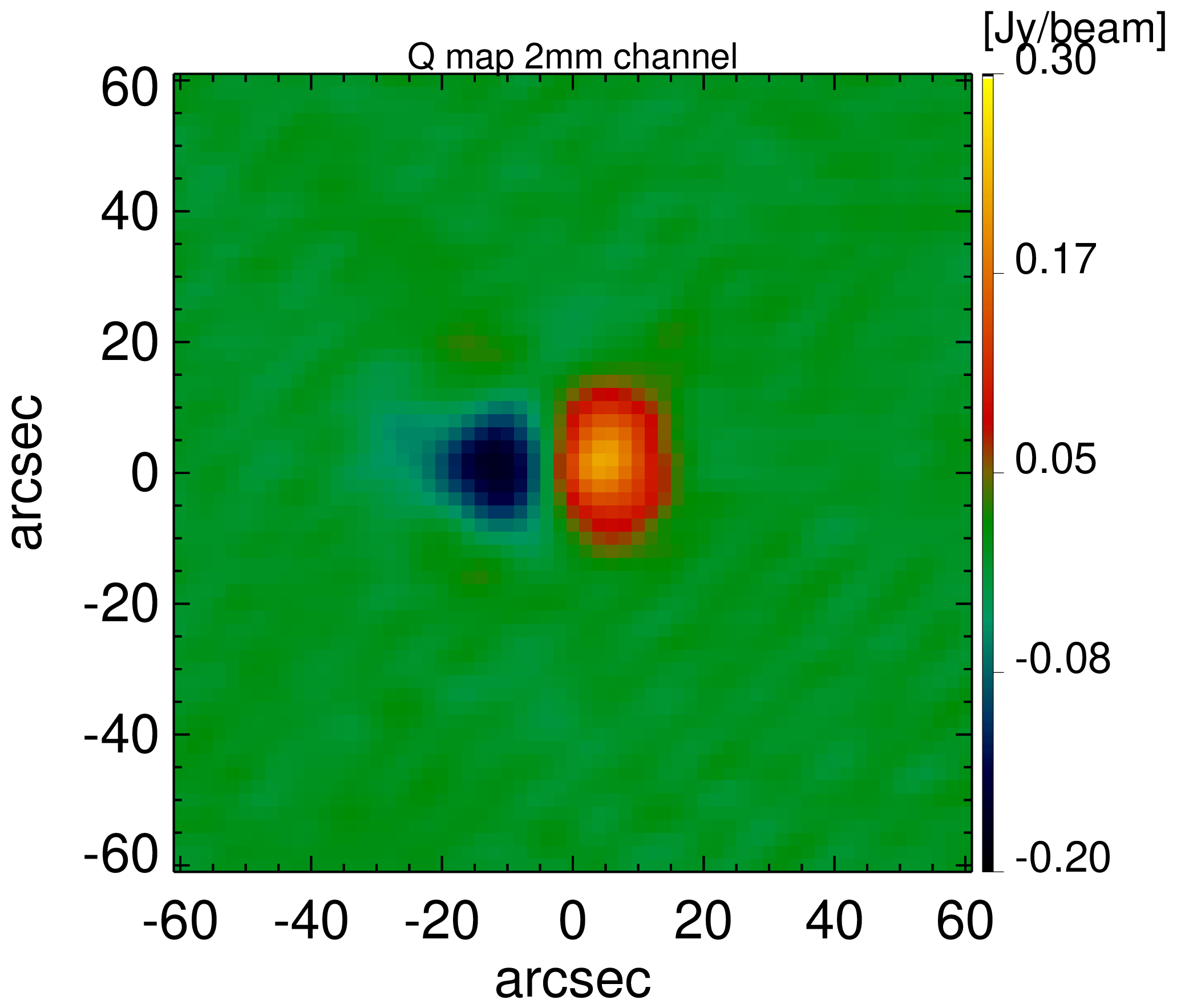}
         \includegraphics[%
               	width=0.33\linewidth,keepaspectratio]{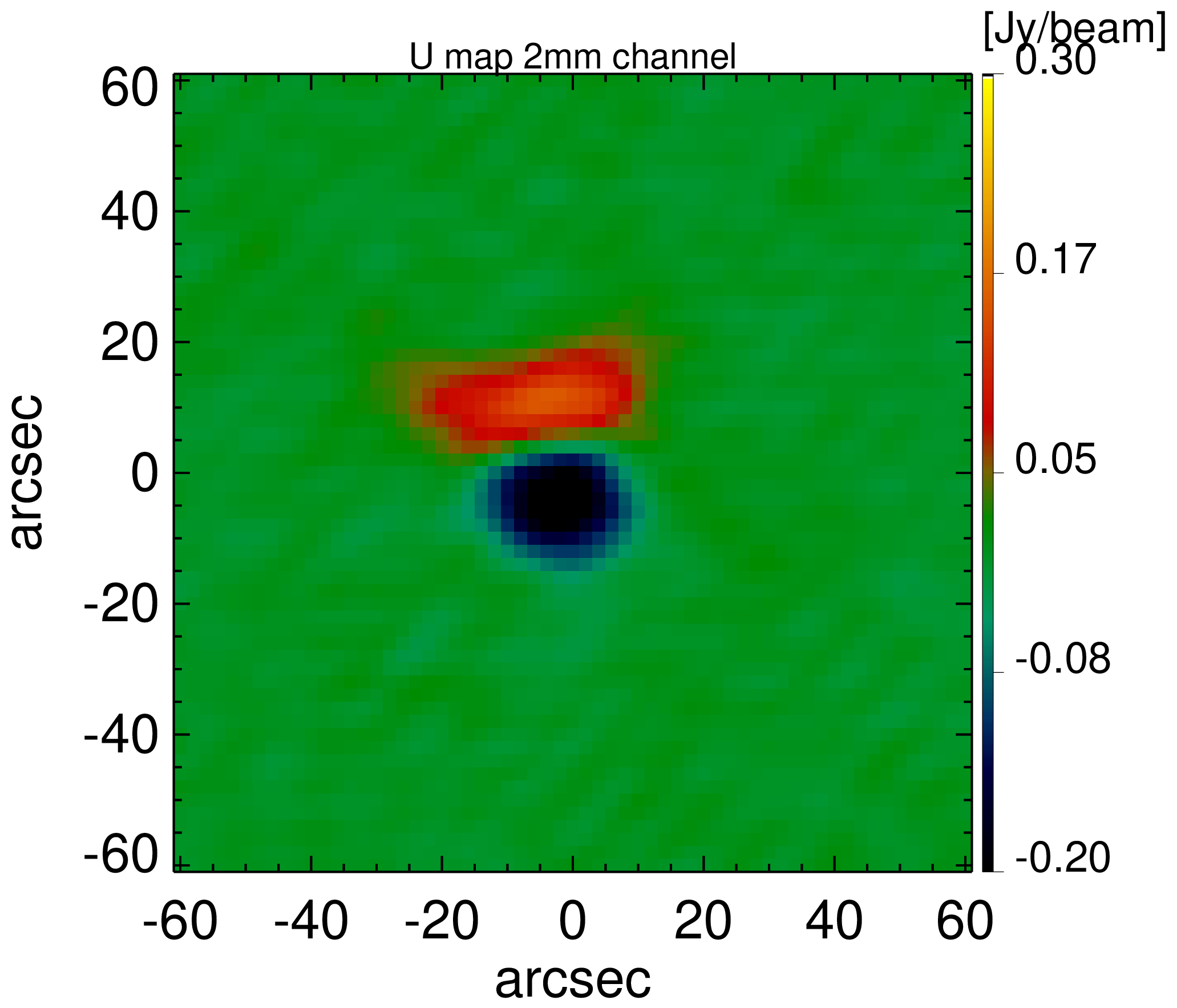}
          \caption{\footnotesize Uranus maps at 2.05 mm. The maps are projected in Nasmyth coordinates} 
          \label{uranus}
 \end{figure}

In order to study the instrumental polarisation, during the last observational campaign on February 2015, we observed the Uranus planet, which is expected to be unpolarised.  The I, Q, U maps observed at 2.05 mm are presented in fig.~\ref{uranus} in Nasmyth coordinates. They show a systematic effect, which consists of a positive and negative feature in the Q and U maps, likely to be due to imperfections on the sub reflectors, a well known problem in radioastronomy  \cite{conway}. 
In terms of polarisation degree the systematic effect corresponds to 2 \%  and 3  \% of the intensity I at 2.05 mm and 1.15 mm, respectively.
We consider that this systematic effect is due to intensity to polarisation leakage. For point sources we can correct for this effect using the observed Uranus intensity to polarisation leakage as a template.
However, for extended sources more complex algorithms need to be developed and this work is in progress. 

\begin{figure}
       \includegraphics[%
       		width=0.33\linewidth,keepaspectratio]{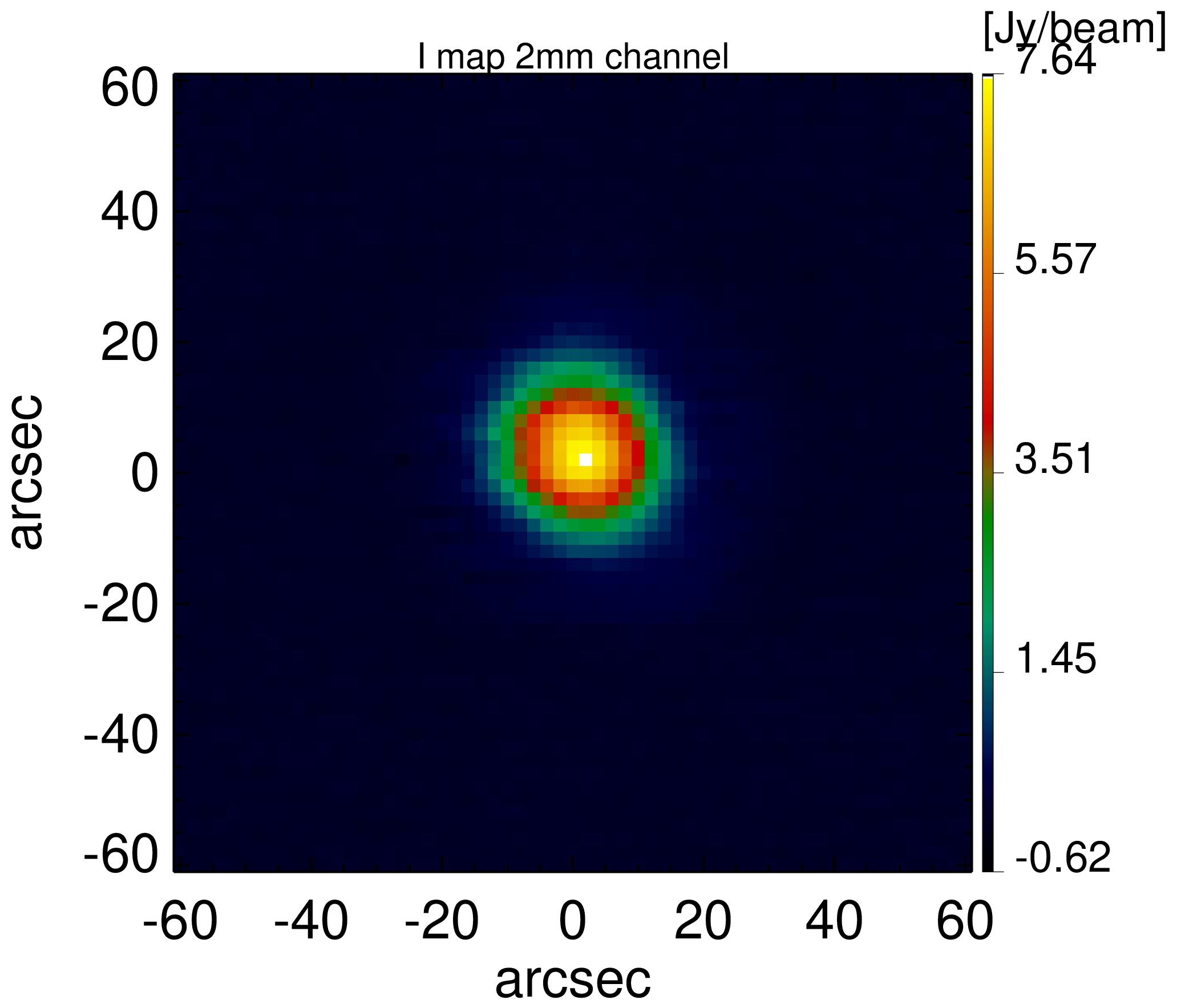}
         \includegraphics[%
         	width=0.33\linewidth,keepaspectratio]{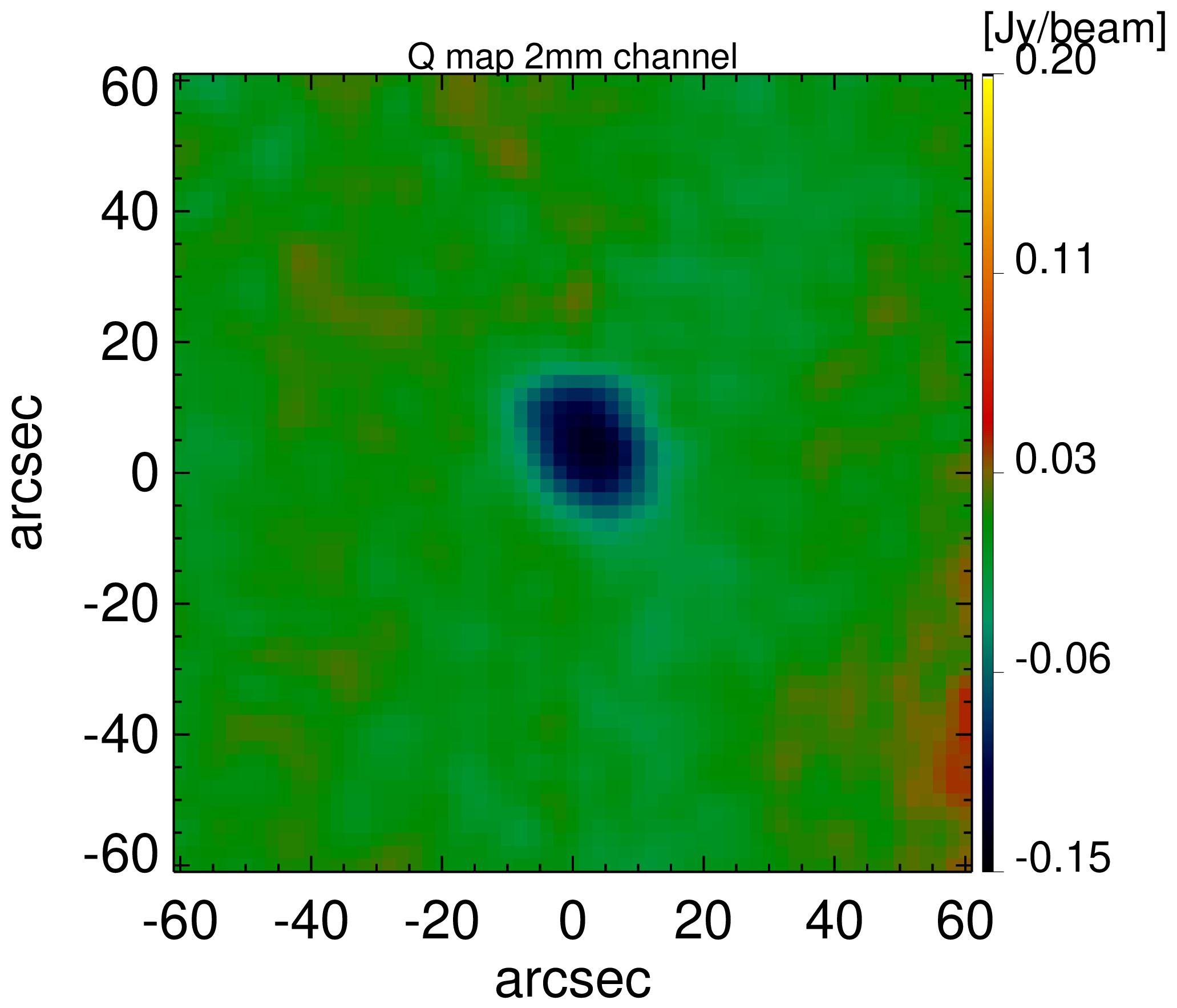}
         \includegraphics[%
         	width=0.33\linewidth,keepaspectratio]{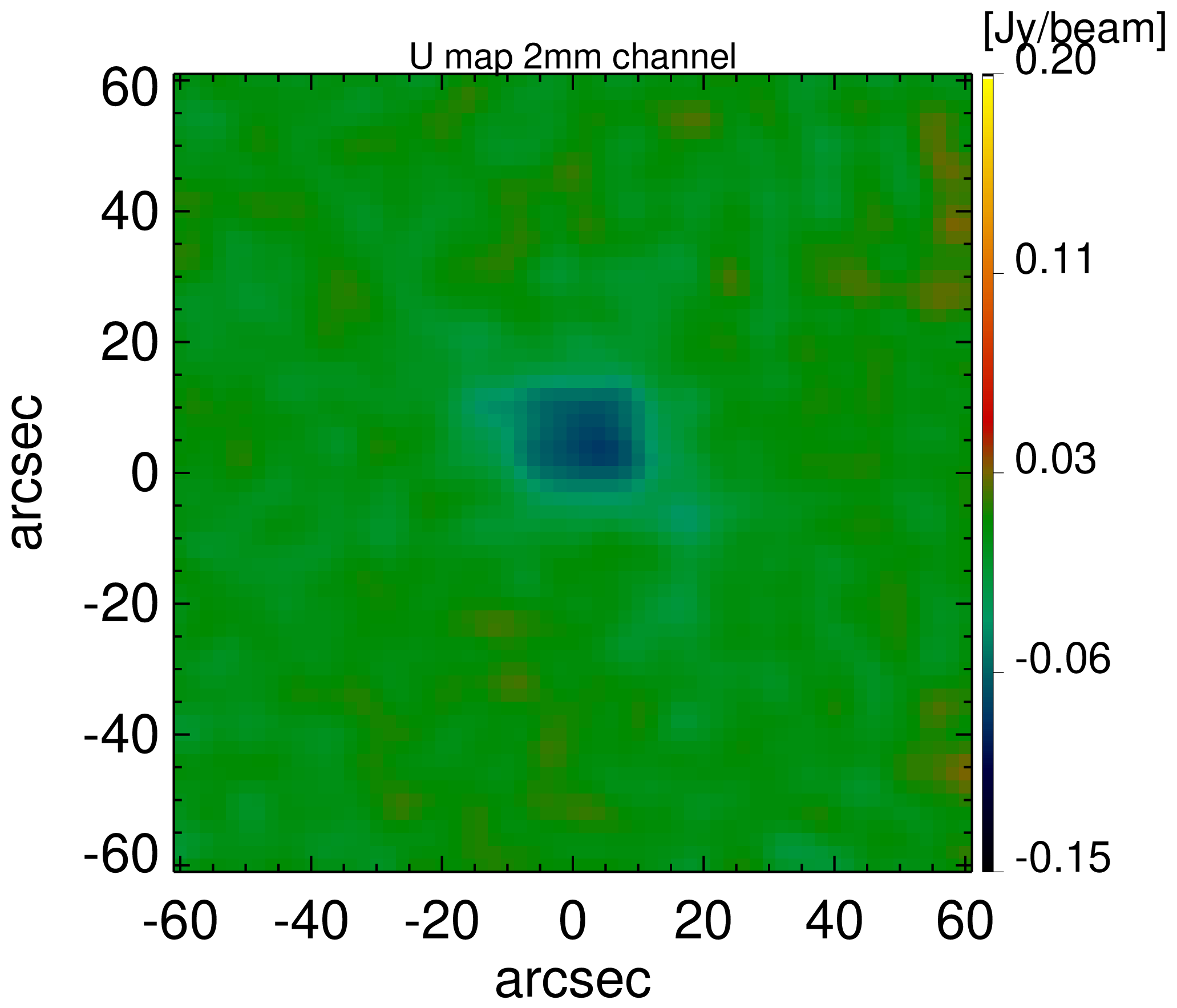}
         \caption{\footnotesize  From left to right I, Q, U maps of 3C273 at 2.05 mm. The maps are corrected for the systematic effect and they are projected in equatorial coordinates.}
\label{3c273}
 \end{figure}

Fig.~\ref{3c273} shows the I, Q, U maps of quasar 3C273 obtained at 2.05 mm after correcting for the systematic effect as discussed above. We obtain a polarisation degree of (3.5 $\pm$ 0.4) \% and (2.1 $\pm$ 0.2) \% at 1.15 mm and 2.05 mm, respectively. For the polarisation angle, defined as ${\psi}$ = $\frac{1}{2} {\rm arctan}\frac{U}{Q}$ we find -88.1{$^\circ$} $\pm$ 3.8{$^\circ$} and -73.8{$^\circ$} $\pm$ 3.1{$^\circ$} at 1.15 mm and 2.05 mm, respectively. These results are consistent with XPol measurements \cite{thum}, which were performed in parallel during the NIKA campaign. 
Additionally, we observed the Crab nebula (Tau A, M1 or NGC 1952) which is a typical polarisation calibrator at mm wavelengths \cite{macias}. 

\begin{figure}
\includegraphics[%
	width=0.33\linewidth,keepaspectratio]{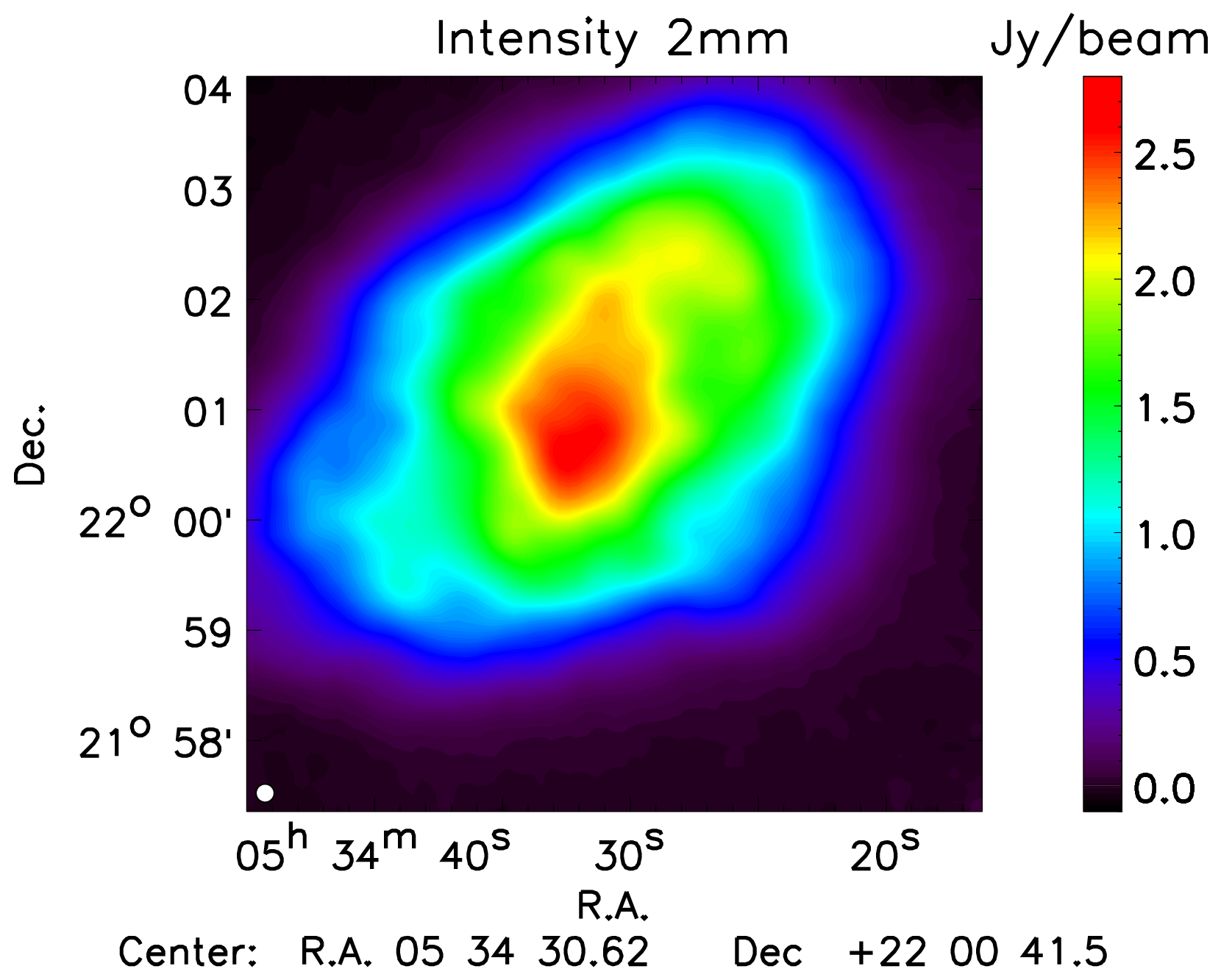}
\includegraphics[%
	width=0.33\linewidth,keepaspectratio]{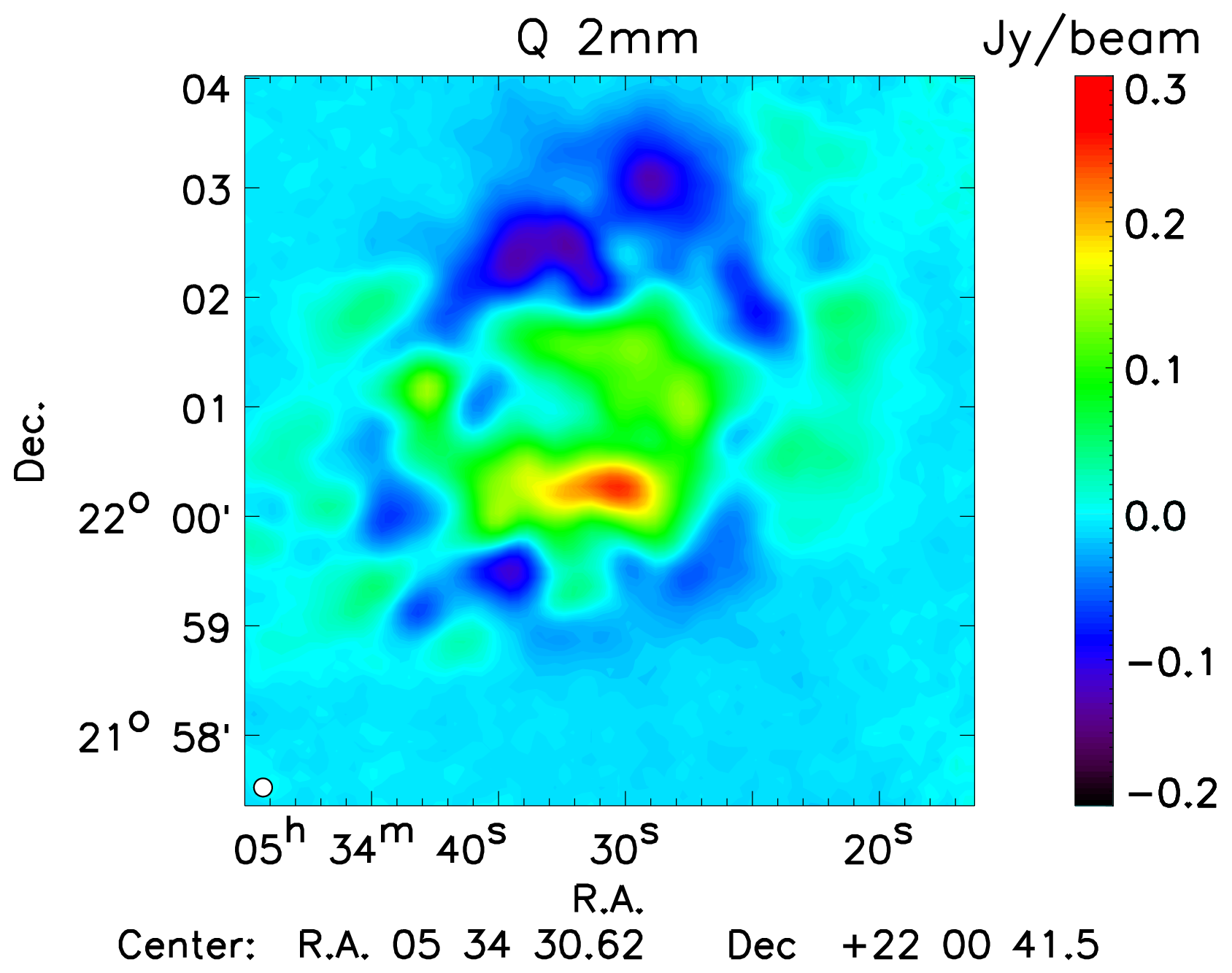}
\includegraphics[%
	width=0.33\linewidth,keepaspectratio]{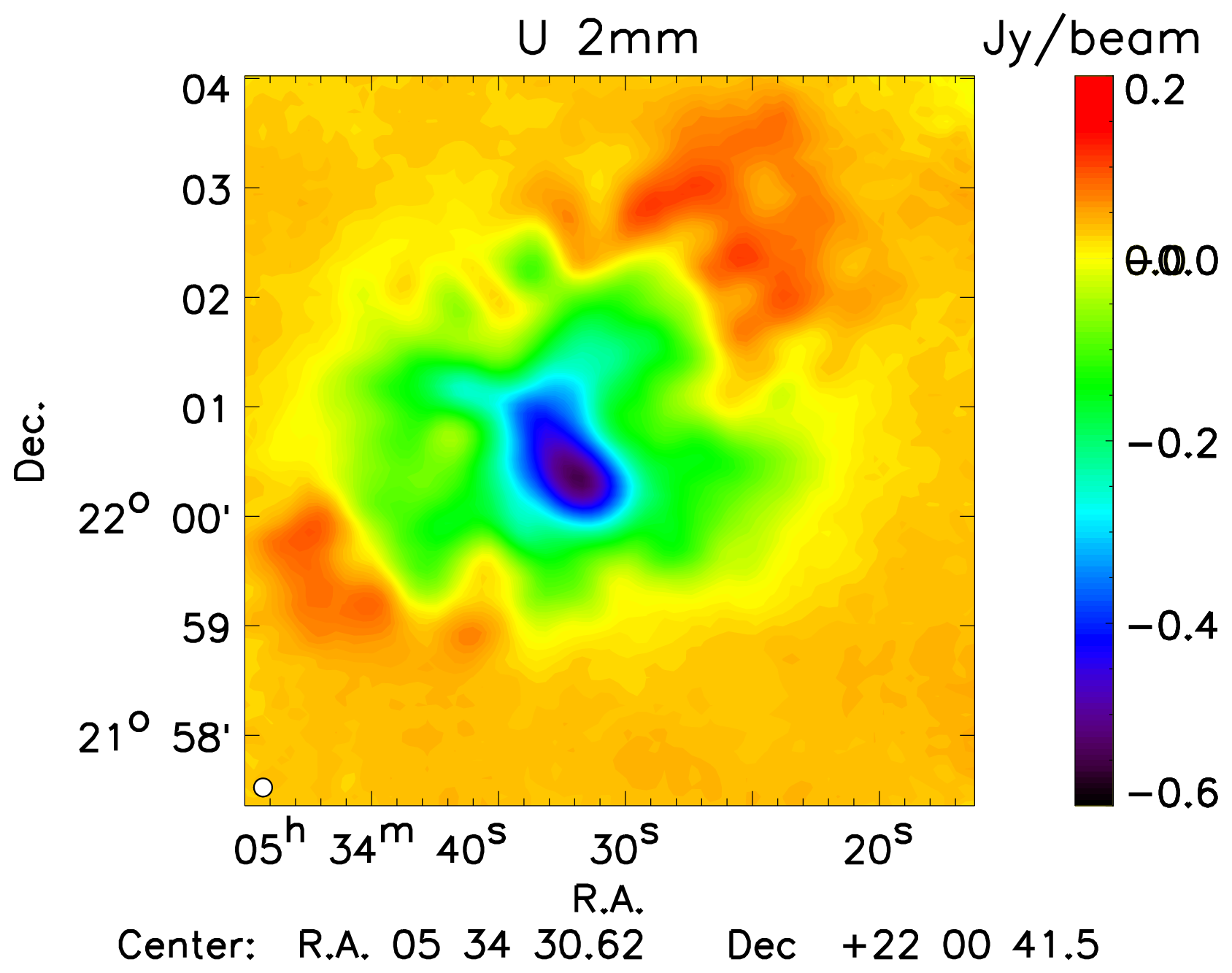}
 \caption{\footnotesize  I, Q, U maps of the Crab nebula at 2.05 mm. (Color figure online)}
 \label{crab}
\end{figure}

The Crab nebula is a supernova remnant that emits a highly polarised signal due both to the synchrotron emission of the central pulsar and its interaction with the surrounding gas. In fig.~\ref{crab} we show the I, Q, U maps of the Crab nebula observed at 2.05 mm without leakage systematic effect correction. Although, the analysis is not fully completed, we observe good qualitative agreement with previous measurements by XPol \cite{aumont}. 
\section{Conclusion}
The optical setup (warm HWP + warm polariser) of the NIKA experiment at the IRAM 30 m telescope has been proved to be efficient for the reconstruction of the sky polarisation. For this first light, a systematic leakage of total intensity into polarisation shows up at the 3$\%$ level. We have been able to correct for it using calibration scan on Uranus. The exact source of this effect is being investigated. 
A similar polarisation setup will be mounted on NIKA2. NIKA2 will have a larger number of detectors than NIKA (5000 instead of 300) and a 100 mK state of the art polariser. Thus, we expect it to be an avant-garde instrument for polarisation measurements. 

\begin{acknowledgements}
We  would  like  to  thank  the  IRAM  staff for  their  support  during  the campaign. This work has been partially funded by the Foundation Nanoscience Grenoble, the ANR under the contracts "MKIDS'' and "NIKA''. This work has been partially supported by the LabEx FOCUS ANR-11-LABX-0013. This work has  benefited  from  the  support  of  the  European  Research  Council  Advanced Grant  ORISTARS  under  the  European  Union's  Seventh  Framework  Program (Grant Agreement no. 291294). The NIKA dilution cryostat was designed and built at the Institut Neel. A. R. would like to thank the FOCUS French LabEx doctoral fellowship program. A. R. acknowledges support from the CNES doctoral fellowship program. R. A. would like to thank the ENIGMASS French LabEx doctoral fellowship program. B. C. acknowledges support from the CNES post-doctoral fellowship program.
\end{acknowledgements}

\end{document}